\documentclass[10pt,showpacs,preprintnumbers,amsmath,eqsecnum,amssymb,twocolumn,nofootinbib,a4paper,superscriptaddress]{revtex4}
\pdfoutput=1

\usepackage{graphicx}

\def\be{\begin{equation}}
\def\ee{\end{equation}}

\def\bea{\begin{eqnarray}}
\def\eea{\end{eqnarray}}

\def\H{{\cal H}}

\def\cssq{c_{\rm{s}}^2}
\def\cesq{c_{\rm{ph}}^2}

\def\U0{{\bar U_0}}

\def\bi{\begin{itemize}}
\def\ei{\end{itemize}}

\usepackage{epsfig}
\usepackage{graphicx,epsf}
\usepackage{color}
\usepackage{bm}
\usepackage{psfrag}

\def\be{\begin{equation}}
\def\ee{\end{equation}}
\def\beb{\begin{equation*}}
\def\eeb{\end{equation*}}
\def\bea{\begin{eqnarray}}
\def\eea{\end{eqnarray}}
\def\beab{\begin{eqnarray*}}
\def\eeab{\end{eqnarray*}}
\def\nn{\nonumber}

\def\H{{\cal H}}

\def\cs2{c_{\rm{s}}^2}

\def \beg {\begin{enumerate}}
\def \en {\end{enumerate}}

\def\cs{c_{\rm{s}}^2}
\def\V1{{\bar{V_1}}}

\begin{document}
\title{Modelling non-dust fluids in cosmology}

\author{Adam J.~Christopherson}
\email[]{Adam.Christopherson@nottingham.ac.uk}
\affiliation{School of Physics and Astronomy, University of Nottingham, University Park,
Nottingham, NG7 2RD, United Kingdom}

\author{Juan~Carlos~Hidalgo}
\email[]{juan.hidalgo@port.ac.uk}
\affiliation{Instituto de Astronom\'ia, UNAM, Ciudad Universitaria, 04510, M\'exico D.F., M\'exico}
\affiliation{Institute of Cosmology and Gravitation, University of Portsmouth,
Dennis Sciama Building, Burnaby Road, Portsmouth, PO1 3FX, United Kingdom}

\author{Karim~A.~Malik}
\email[]{k.malik@qmul.ac.uk}
\affiliation{Astronomy Unit, School of Physics and Astronomy, Queen Mary University of London,
Mile End Road, London, E1 4NS, United Kingdom}

\date{\today}

\begin{abstract}
Currently, most of the numerical
  simulations of structure formation use
  Newtonian gravity. When modelling pressureless dark matter, or `dust', this
approach gives the correct results for scales much smaller than
    the cosmological horizon, but for scenarios in which the
fluid has pressure this is no longer the case.  In this article, we
present the correspondence of perturbations in Newtonian and cosmological
  perturbation theory, showing exact mathematical 
equivalence for pressureless matter, and giving the relativistic
corrections for matter with pressure. As an example, we study
    the case of scalar field dark matter which 
features non-zero pressure perturbations. We discuss some problems
which may arise when evolving the
    perturbations in this model with Newtonian numerical simulations
and with CMB Boltzmann codes. 
\end{abstract}

\pacs{98.80.Cq \hfill  arXiv:1207.1870}

\maketitle

\section{Introduction}

Current observations indicate that the universe in which we live is
almost homogeneous and isotropic. However, it is also known that
small initial departures from homogeneity and isotropy give rise to the
structures we observe today. Thus, while the universe is well 
approximated on large scales by a homogeneous and isotropic
Friedmann-Lema\^itre-Robertson-Walker (FLRW) spacetime, the existence
of large scale structure and inhomogeneities in the Cosmic Microwave
Background (CMB) tell us that this is not the complete picture \cite{llbook}.

In order to model inhomogeneities we utilise a technique that is well
established in many branches of physics and applied mathematics,
namely, we take an approximate solution and add small
perturbations. In cosmology, this technique is
called cosmological perturbation theory, and requires the addition of
small inhomogeneous perturbations on top of the FLRW background, such
that the system still solves Einstein's field equations
(e.g. Refs.~\cite{ks, mfb,  MM2008}). When considering the dynamics on 
sufficiently small scales, and for fluids which are pressureless, it
is enough to use Newtonian physics
\cite{Peebles}. Inhomogeneous perturbations of Newtonian cosmology
have been studied for many years and these are the equations that are
used when performing large numerical simulations of galaxy formation
\cite{Bertschinger:review,Bernardeau:review}.  However, as simulations get more
sophisticated and the observations more precise, it is possible to
    distinguish those components which exhibit pressure
perturbations. For example, to fully account for the effects of
inhomogeneous scalar fields as dark energy
(e.g. \cite{Hu:1998tj,Ma:1999dwa,Garriga:1999vw,ArkaniHamed:2003uy}),
dark matter \cite{Peebles:1999fz,Sahni:1999qe,Hu:2000ke,Matos:2001ps},
or a unified field
\cite{Scherrer:2004au,Liddle:2006qz,Bertacca:2007ux,DeSantiago:2011qb}, 
in the formation of structures, one should employ more general
equations with the input of general relativity.    

In this paper we re-visit the question of how to relate Newtonian and
cosmological perturbation theories\footnote{Note, throughout this
  paper we work within the confines of linear perturbation
  theory.}. We show that, for pressureless systems, the equations
governing cosmological perturbations of an FLRW spacetime reduce to
the equivalent Newtonian equations on using gauge invariant variables
-- the metric potential in the longitudinal gauge, and the density
contrast in the comoving gauge or the total matter gauge.  Drawing on
this equivalence, we then investigate the situation for fluids with
pressure and/or pressure perturbations. We find that one can write the
Poisson equation in the usual way, but that the continuity and Euler
equations then differ, depending upon the equation of state parameter
and the adiabatic sound speed.

We then go on to study how to relate the two perturbation theories in
a general scalar field model. We find that,
as expected, the Poisson equation is identical to
the Newtonian case, but that the Euler and continuity equations
differ, depending now on the equation of state parameter and the
effective speed of propagation of perturbations through the system.
Finally we discuss the Jeans scale in the scalar field dark matter
models, where the background equation of state parameter is zero. We
find that this scale depends upon $c_{\rm ph}^2$, the phase speed, or
speed of propagation of perturbations. We close, in
Section~\ref{sec:dis}, with a brief discussion.

\section{Modelling Inhomogeneities}

\subsection{Newtonian perturbations}

Let us first study the theory of perturbations in Newtonian physics.
We consider inhomogeneous perturbations about a 
homogeneous background, and so the energy density $\rho$ is
\be 
\rho(\vec{x},t)=\bar{\rho}(t)\Big(1+\delta_{\rm {N}}(\vec{x},t)\Big)\,,
\ee
where $\bar{\rho}(t)$ is the homogeneous background energy density and
$\delta_{\rm {N}}(\vec{x},t)$ is the inhomogeneous density contrast. On
introducing the inhomogeneous Newtonian potential, $\Phi_{\rm
  N}(\vec{x},t)$ and fluid velocity $\vec{v}_{\rm {N}}(\vec{x}, t)$, the
linearised conservation and Euler equations are, respectively,
\cite{Peebles, White} 
\begin{align}
&\dot{\delta}_{\rm {N}}+\frac{1}{a}\vec{\nabla}\cdot\vec{v}_{\rm {N}}=0\,,
\label{newtonian:cons}\\
&\dot{\vec{v}}_{\rm {N}}+H\vec{v}_{\rm {N}}=-\frac{1}{a}\vec{\nabla}\Phi_{\rm
  N}-\frac{1}{a\bar{\rho}}\vec{\nabla}\delta P_{\rm {N}}\,,
\label{newtonian:euler}
\end{align}
where $H=\dot{a}/a$ is the Hubble parameter, a dot denotes a
derivative with respect to coordinate time and $\delta P_{\rm
 {N}}$ denotes the pressure perturbation. 
The Newtonian potential and the density contrast are then related
through the Poisson equation  
\be 
\label{eq:Npoisson}
\nabla^2\Phi_{\rm N}=4\pi G a^2 \bar{\rho}\delta_{\rm {N}}\,,
\ee
where the Laplacian is defined as $\nabla^2=\vec{\nabla}\cdot\vec{\nabla}$.
On utilising the relationship between the energy density perturbation
and the pressure perturbation for a barotropic fluid, $\delta P =
c_{\rm s}^2 \delta\rho$, we can combine the fluid equations into a
second order equation
\be 
\label{eq:delta2nd}
\frac{\partial^2 \delta_{\rm N}}{\partial t^2}+2H\frac{\partial
  \delta_{\rm N}}{\partial t}=4\pi G\bar{\rho}\delta_{\rm N} 
+c_{\rm s}^2\nabla^2\delta_{\rm N}\,.
\ee
\subsection{Cosmological perturbations}

While the theory of Newtonian perturbations is sufficient for
modelling small scale  
physics involving only pressureless { {dust}} particles, 
the dynamics of the universe are governed by general
relativity. Therefore, we must consider relativistic perturbation
theory. Since Einstein's theory relates the geometry of the universe
to its matter content, we must consider perturbations of both the
matter and the FLRW spacetime metric.   

The most general, linear scalar perturbations to the FLRW metric are
\cite{mfb, ks, MW2008} 
\begin{align} 
ds^2=a^2(\eta)\Big[&-(1+2\phi)d\eta^2+2B_{,i} dx^i d\eta \nn\\
&+\{(1-2\psi)\delta_{ij}+E_{,ij}\}dx^idx^j\Big]\,,
\end{align}
where we now use conformal time $\eta$, related to coordinate time $t$
through $dt=ad\eta$. A unique problem which arises in the relativistic
theory is the problem of gauge dependence. Since general relativity is
covariant, and splitting 
the spacetime into a background and a perturbation is not a covariant
process, we introduce extra spurious coordinate dependence (see, e.g.,
Refs.~\cite{stewart1990, MM2008, Malik:2012dr}). This can be resolved in a
systematic manner, as was first shown by Bardeen
\cite{Bardeen:1980kt}, by considering gauge-invariant variables --
quantities that do not change under a gauge transformation. A popular
choice of variables amount to setting $E$ and $B$ to zero, resulting
in the FLRW metric in the so-called longitudinal or Newtonian gauge,
with the gauge invariant variables $\Phi$ and $\Psi$
\be 
ds^2=a^2(\eta)\Big[-(1+2\Phi)d\eta^2+(1-2\Psi)\delta_{ij}dx^idx^j\Big]\,.
\ee
The conservation and Euler equations are then reduced from
the form without fixing the gauge   
\begin{align}
& \delta \rho' + 3 \H (\delta \rho + \delta P) - 3(\bar{\rho} +
  \bar{P})\psi'\nn\\
& \qquad\qquad\qquad + (\bar{\rho} + \bar{P} )\nabla^2 (v + E') = 0\,,
\label{gauge:cont}\\
&V' + \Big(1 -3 \frac{\bar{P}'}{\bar{\rho}'}\Big) \H V + \phi + 
\frac{1}{\bar{\rho} + \bar{P}} \left(\delta P +\frac{2}{3} \nabla^2 \Pi \right) = 0\,,
\label{gauge:euler}
\end{align}
to the following expressions, where we neglect anisotropic stresses, $\Pi$,
\begin{align}
\label{eq:cons_rel}
&\delta\rho_{\ell}'+(\bar{\rho} + \bar{P})(\nabla^2 v_{\ell}-3\Psi')=
- 3\H(\delta 
\rho_{\ell} + \delta P_\ell)\,,\\
\label{eq:euler_rel}
&\left[(\bar{\rho} + \bar{P} )v_{\ell}\right]'+ 4\H(\bar{\rho}+\bar{P})v_{\ell}+
(\bar{\rho}+\bar{P})\Phi+ \delta P_{\ell}=0\,.
\end{align}
%
%
%
Here the subscript $\ell$ denotes matter variables in the
longitudinal gauge; $v$ is the velocity potential, i.e. $v^i =
\nabla^i v$ with $v^i$ the fluid three velocity; and the
    momentum potential is $V\equiv v+B$.  

We can then specialise to a barotropic fluid with equation of state
$\bar{P}=w(\eta)\bar{\rho}$, and whose pressure perturbation can be
related to the energy density perturbation through 
\be 
\label{eq:dP}
\delta P=c_{\rm s}^2\delta\rho\,,
 \ee
where $c_{\rm s}^2=\bar{P}' / \bar{\rho}'$ is the adiabatic sound
speed\footnote{The adiabatic sound speed, equation of state parameter
  and its derivative are related through 
\be 
\frac{w'}{1+w}=-3\H(c_{\rm s}^2-w)\,.
\ee
For the case where $w$ is constant, this relationship
guarantees that $w=c_{\rm s}^2$.}. Thus, for this system,
Eqs.~(\ref{eq:cons_rel}) and (\ref{eq:euler_rel}) become 
\begin{align}
&\delta_{\ell}'+(1+w)(\nabla^2 v_{\ell}-3\Psi')=3\H(w-c_{\rm s}^2)\delta_{\ell}\,,\\
&v_{\ell}'+\H(1-3 c_{\rm s}^2 )v_{\ell}+\Phi+\frac{c_{\rm s}^2}{1+w}\delta_{\ell}=0\,.
\end{align}


The Einstein equations then give that $\Phi=\Psi$ (in the case of zero
anisotropic stress, as is true for any perfect fluid), and the Poisson
equation is  
\be 
\nabla^2\Psi=4\pi G a^2\bar{\rho}\Big[\delta_{\ell}-3\H(1+w) v_{\ell}\Big]\,.
\ee
In order to draw an equivalence between this and the Newtonian Poisson
equation, we consider
that the energy density perturbation transforms under the gauge
transformation $x^\mu \to \widetilde{x^\mu}\equiv x^\mu+\delta\eta$ as
\be 
\widetilde{\delta\rho}=\delta\rho+\bar{\rho}'\delta\eta\,,
\ee
we get for the comoving density contrast, in terms of the 
longitudinal density contrast and the velocity perturbation \cite{MW2008},
\be 
\label{eq:contrast_com}
\delta_{\rm c}=\delta_{\ell}-3\H(1+w)v_{\ell}\,.
\ee
Then we obtain the Poisson equation \cite{Bardeen:1980kt,Wands:2009ex}
\be 
\nabla^2{{\Psi}}=4\pi G a^2 \bar{\rho}\delta_{\rm c}\,,
\ee
which is equivalent to Eq.~(\ref{eq:Npoisson}) upon the
identification 
\be
\Psi =  \Phi_{\rm N}\,, \qquad \delta_{\rm c} = \delta_{\rm N}\,.
\ee

\noindent This equivalence is the one we follow in the rest of the
paper. Historically, this is the reason why the longitudinal gauge is
at times called the Newtonian gauge.  

\subsection{Correspondences}

In this Section we relate the relativistic equations to the
Newtonian equations. We take heed from the Poisson equation which, as
stated above, relates the density contrast in the comoving gauge to
the metric potential in the longitudinal gauge:
\be
\nabla^2\Psi=4\pi G a^2 \bar{\rho}\delta_{\rm c}\,.
\ee
In order to obtain an equivalence for the set of equations, we must
first ensure that we are consistent with the density contrast that we
use. With the aid of the background Friedmann equations and one of
Einstein's perturbed equations, 
\be 
\Psi'=-\H\Psi-4\pi Ga^2(1+w)\bar{\rho}v_\ell\,,
\ee
we can rewrite Eq.~(\ref{eq:cons_rel}) as
\begin{align}
\delta_{\ell}'+3\H &(c_{\rm s}^2-w)\delta_{\ell} + (1+w)\Big(3\H \Psi
  +\nabla^2 v_{\ell}\Big)
\nn\\
\label{eq:cons_1}
&+\frac{9}{2}\H^2 (1 + w)^2v_{\ell}=0\,.
\end{align}
Using now the relationship between $\delta_\ell$ and $\delta_{\rm c}$,
Eq.~\eqref{eq:contrast_com},  
and its derivatives, the continuity equation can be written as
\be 
\label{eq:cont_com}
\delta_{\rm c}'-3\H w\delta_{\rm c}+(1+w)\nabla^2v_{\ell} =0\,.
\ee
Furthermore, the Euler equation, (\ref{eq:euler_rel}), is
\be 
\label{eq:euler_com}
v_{\ell}'+\H v_{\ell} = -\Psi - \frac{\cssq}{1+w} \delta_{\rm c}\,.
\ee

From this we can see that, in the case of a pressureless dust for
which $w=c_{\rm s}^2=0$, the evolution equations reduce to 
\begin{align}
&\delta_{\rm c}'+\nabla^2v_{\ell}=0\,,\\
&v_{\ell}'+\H v_{\ell}+\Psi=0\,.
\end{align}
\noindent These equations are finally identical to the
    Newtonian \eqref{newtonian:cons} and
    \eqref{newtonian:euler}. Thus we can establish a
    mathematical equivalence between Newtonian and relativistic
    velocity for the  case of dust, 
    \be
   v_{\ell} = v_{\rm N}.
    \ee

\noindent The above equations can be recast as a second order
differential equation for the density contrast as   
\be 
\label{eq:deltasecond}
\delta_{\rm c}''+\H \delta_{\rm c}'=4\pi Ga^2\bar{\rho}\delta_{\rm c}\,,
\ee
\noindent which governs the evolution of matter density
perturbations. Here the second term is a suppression of the
perturbations with  the expansion of the universe, and the term on the
right hand side sources the growth of perturbations due to the gravitational
instability.  
This is the equivalent form to the Newtonian perturbation theory (as
expected for a non-relativistic species), as obtained in
Eq.~(\ref{eq:delta2nd}). 

The exact mathematical equivalence between the set of variables
    presented and their Newtonian counterpart, is already known
    and has been presented for the Poisson and the evolution equations
    in the context of other problems
    \cite{Hu:2004xd,Wands:2009ex,Chisari:2011iq}.  

However, should one wish to consider fluids other than dust, with
non-zero pressure either in the background or in the perturbations,
then the relativistic equations must be used instead. In such cases,
the exact mathematical equivalence with the Newtonian counterpart does
not hold and the form of the equations in numerical studies must
change.   

This is especially important for, say, hot dark
matter, or for a system containing dark energy perturbations.
In the  following, we will consider the case of scalar field models of
dark matter.   
 
\section{Scalar fields as perfect fluids}
\label{sec:SFDM}

In this section we aim to present the minimal additions to the
ordinary hydrodynamic equations to treat cosmological structure
formation when the matter components support pressure. Specifically,
models of scalar fields treated as fluids with zero effective
pressure, yet with non-negligible pressure perturbations, have been
considered to account for the dark matter component
\cite{Hu:1998kj,Bertacca:2007ux,Matos:2000ss}.  

The basic model features a canonical scalar field oscillating at the
bottom of its potential with a period much smaller than the Hubble
time and any other dynamical times. A model of dark matter which has
received increased attention is the scalar field dark matter model
(SFDM), specifically in the form of a Bose-Einstein condensate
\cite{UrenaLopez:2008zh,Lundgren:2010sp}, the theory is robust enough
to study the numerical problem of structure formation (recent attempts
are \cite{Woo:2008nn, suarez:structure,Magana:2012xe}). In contrast to
cold dark matter (CDM), this component exhibits pressure
perturbations\footnote{ There 
      are conversely, models which unify dark energy and dark
      matter under a single degree of freedom
      \cite[e.g.][]{Lim:2010yk,Aviles:2011ak}. The set of equations in
      the previous section are sufficient to work with in these cases,
      with the equation of state as the only degree of freedom.}.   

It is well known that a scalar field system cannot be modelled as a
barotropic fluid (except on super-horizon scales \cite{nonad}), and
in fact a consistent fluid equivalence must consider a more general
perfect fluid~\cite{mainini:08}. Indeed, as elucidated in
Ref.~\cite{Valiviita:2008iv}, if one is to
interpret a scalar field as 
a fluid, a distinction must be made between
 $w$ and $c_{\rm s}^2$. Furthermore, the speed with which pressure
perturbations propagate is described by the effective sound speed, or
phase speed, $c_{\rm ph}^2$ defined, in the fluid rest frame, as 
\be 
\label{eq:ceff}
c_{\rm ph}^2=\frac{\delta P}{\delta \rho}\Bigg|_{\rm rf}\,.
\ee
For a scalar field with a canonical kinetic term, this is equal to unity.
However, for a non-canonical scalar field with pressure and energy density
depending upon both the field, $\varphi$, and its kinetic term, 
$X\equiv-\frac{1}{2}g^{\mu\nu}\partial_\mu\varphi\partial_\nu\varphi$,
this can differ from one and is given by \cite{Garriga:1999vw,
  nonad}\footnote{Note that the equivalence between
  Eqs.~(\ref{eq:ceff}) and (\ref{eq:cph}) is only true for a single
  scalar field, which is the case we study in this paper.}
\be 
\label{eq:cph}
c_{\rm ph}^2=\frac{P_{,X}}{\rho_{,X}}\,.
\ee
In general, for a scalar field, the pressure
perturbation is no longer proportional to the energy density
perturbation. Instead the relationship between
pressure perturbations and energy density perturbations in an
unspecified gauge is    
\be 
\delta P = c_{\rm s}^2\delta\rho + (c_{\rm ph}^2-c_{\rm
  s}^2)\Big[\delta\rho+\bar{\rho}'(v+B)\Big]\,. 
\ee
\noindent The second term on the right hand side is often referred to as the
non-adiabatic pressure perturbation $\delta P_{\rm nad}$. 
In this case, Eqs.~(\ref{eq:cons_rel}) and (\ref{eq:euler_rel}) become
\begin{align}
&\delta_{\ell}'+(1+w)(\nabla^2 v_{\ell}-3\Psi')\nn\\
&\qquad=3\H(w-c_{\rm ph}^2)\delta_{\ell}
+9\H^2(1+w)(c_{\rm ph}^2-c_{\rm s}^2)v_\ell\,,\\
&v_{\ell}'+\H(1-3w)v_{\ell}+\Phi+\frac{c_{\rm ph}^2}{1+w}\delta_{\ell}+\frac{w'}{1+w}v_\ell\nn\\
&\qquad=3\H(c_{\rm ph}^2-c_{\rm s}^2)v_\ell\,.
\end{align}

As shown in the previous section, in order to write the equations in
the variables employed in simulations, we present the equations in
longitudinal gauge and then introduce the density
contrast in the comoving gauge. Following a similar procedure, using
the background and perturbed Einstein equations, we have
\begin{align}
&\delta_{\ell}'+3\H(\cesq-w)\delta_{\ell} + 3\H^2(1+w)\Psi
  +(1+w)\nabla^2 v_{\ell}
\nn\\
&+9\H^2(1 + w)(\cssq - \cesq+ \frac{1+w}{2})v_{\ell}=0\,,
\end{align}
\noindent which in light of Eq.~\eqref{eq:contrast_com} can then be written as  
\be 
\delta_{\rm c}'-3\H w\delta_{\rm c}+(1+w)\nabla^2v_{\ell} =0\,.
\label{continuity:perfect}
\ee
{Interestingly,} this is identical to the equation for the
barotropic fluid presented earlier.  
However, the Euler equation does not take the 
same form as Eq.~(\ref{eq:euler_com}). Instead, in terms of
the comoving density contrast, it is 
\be 
v_{\ell}'+\H v_{\ell} = -\Psi - \frac{\cesq}{1+w} \delta_{\rm c}\,.
\label{euler:perfect}
\ee

\noindent This form shows the advantage of
    using the same variables that are identified with the Newtonian
    counterpart in the dust case. These last two
    equations present explicitly the minimal relativistic modification
    with respect to the dust case
    in equations \eqref{newtonian:cons} and
    \eqref{newtonian:euler}.
  
The combination of these two equations yields a Klein-Gordon
equation -- a generalisation of Eq.~(\ref{eq:deltasecond})
for $\delta$ with an extra Laplacian term,
\begin{align}
\delta_{\rm c}''  - c_{\rm ph}^2 \nabla^2 \delta_{\rm c} & + \H
\left(6w-3c_{\rm s}^2 - 1 \right)\delta_{\rm c}'
\notag\\
& - 3 \H^2 \left(\frac{1}{2} + 4 w - \frac{3}{2} w^2 - 3 c_{\rm s}^2  \right)
\delta_{\rm c} = 0.  
\label{perfect:kg}
\end{align}

\noindent The above equation is reduced to
    Eq.~\eqref{eq:deltasecond} in the limit when $w  = c_{\rm s}^2 =
    c_{\rm ph}^2 = 0 $.

As a characteristic feature of the problem one can determine,
directly from Eq.~\eqref{perfect:kg} the instability scale 
for density perturbations. For the most general case we
obtain the Jeans wavenumber  
\be
k_{\rm J}^2 = \frac{3\ \H^2}{2\ c_{\rm ph}^2}\left( 1 + 8 w - {3} w^2 -
6 c_{\rm s}^2 \right) \,. 
\label{perfect:jeans}
\ee
\noindent This scale is associated to a power spectrum cutoff, and is a
characteristic feature of non-dust components. In the case of the
scalar field acting as a dark matter component, for which $w_{\rm eff} = 0$,
the Jeans' wavelength reduces to
\be
\lambda_{\rm J} = c_{\rm ph} \sqrt{\frac{\pi}{G \bar{\rho} (1 - 6 c_{\rm s}^2)}}.
\ee

\noindent This is consistent with the fact that the speed of propagation of
perturbations is given by $c_{\rm ph}^2$.  

The effective sound speed for each canonical SFDM model is equivalent
and equal to 1, though for non-canonical models this differs (see
e.g. \cite{Bertacca:2008uf}). The specific scales for each model, as
well as the growth of perturbations will be treated elsewhere. For our
purposes it suffices to note that the $c_{\rm s}^2$ contribution to
the last equation would not be manifest in the Newtonian context. This
is eventually important in determining the scale of a spectrum cutoff
\cite{Hu:1998kj,Matos:2000ss}. 

These simple results show the importance of considering the system
with no approximations and argue for the use of the system of
equations \eqref{continuity:perfect} and \eqref{euler:perfect} in the
forthcoming simulations of structure formation in the general SFDM models. 

Another important  comment is that care must be taken when
studying the SFDM model using CMB codes. 
Many of the popular Boltzmann codes are written in the synchronous
gauge, a gauge specified by demanding that $\phi=0=B$. However, as has 
been known for some time, these conditions alone do not fix the gauge. That is,
the synchronous conditions alone are not a complete gauge choice,
and one needs an additional condition. The condition usually employed,
e.g. by CAMB \cite{Lewis:1999bs}, is to set the velocity perturbation
of the cold dark matter to zero.  

This can be done since the dark matter is assumed to have zero
pressure perturbations \cite{thesis}. 
But $v_{\rm DM} = 0$ is a gauge choice that does not admit pressure
perturbations in the dark matter component, since imposing this
condition will result in inconsistencies in the theory (c.f. equation
\eqref{gauge:euler}).
Consequently, in order to use CMB codes to study the SFDM model,
one must either include an extra CDM component to allow for this
condition -- which arguably reduces the value of the theory -- or use a
code that is not written in the synchronous gauge
\citep[e.g.][]{Lesgourgues:2011re}.

\section{Discussion}
\label{sec:dis}

In this paper we have revisited the issue of relating Newtonian
perturbation theory to relativistic perturbation theory in
cosmology. After reviewing both
perturbation theories, we explicitly showed how one relates
them for the case of a pressureless fluid, such as dark
matter. As is well documented in the literature on cosmological
perturbation theory, in order to relate the two approaches one must
use the Newtonian gauge with the comoving energy density
perturbation. With this gauge choice, we have shown that the
hydrodynamical and Poisson equations remain identical to the Newtonian
prescription.  
 
We then considered the extension of this for fluids with pressure,
both for the barotropic and for the perfect fluids taking our lead
from the dark matter case. We show that the mathematical equivalence
is lost and, instead, the continuity equation differs, depending upon
the equation of state parameter, and the Euler equation depends upon
$c_{\rm s}^2$.  

The major application that we explored in Section~\ref{sec:SFDM}
regards scalar field dark matter.  In this case, the dark matter
species is no longer pressureless, but is instead a scalar field with
canonical kinetic term. In addition to the adiabatic sound speed and
equation of state parameter, a scalar field also has a speed of
propagation of perturbations, which we dubbed the phase, or effective
sound speed $c_{\rm ph}^2$. The evolution equations for this scalar field
fluid in the longitudinal gauge depend on all the parameters. However,
on writing the equations in terms of the comoving energy density
perturbation, the Poisson and continuity equations reduce to those of
perfect fluid form. The Euler equation, on the other hand, still
depends upon all the parameters. 

Thus we have shown that, when studying the scalar field dark matter
model and treating it as a fluid, one cannot simply use the
pressureless cold dark matter or the barotropic fluid equations
without potentially finding erroneous results. Instead, one must use
the equations given in Section~\ref{sec:SFDM}. These conclusions can
be extended to quintessence models as well as those modified gravity
theories conformally equivalent to a scalar field.

The inequivalence between both theories is also important in systems
containing more than one fluid. For example, as shown in
Ref.~\cite{Christopherson2010}, in a system containing normal CDM and
a dark energy component, Eq.~(\ref{eq:deltasecond}) is no longer
satisfied, since one must take into account the dark energy
perturbations. \\

Finally we would like to stress again that the results presented here
have been limited to linear or first order, both for Newtonian theory
as well as for cosmological perturbation theory. Linear theory is only
an approximation, as General Relativity is non-linear. We will extend
the results presented here to second order in perturbation theory in
future work \cite{HCM:2012}.

\section*{Acknowledgements}
AJC acknowledges support from the European Commission's Framework
Programme 7, through the Marie Curie International Research Staff
Exchange Scheme LACEGAL (PIRSES-GA​-2010-2692​64) and is grateful to the
IA and ICN, UNAM for hospitality. The authors acknowledge the support
of the DGAPA-UNAM through the grant PAPIIT IN116210-3.  
AJC is funded by the Sir Norman Lockyer Fellowship of the Royal
Astronomical Society, JCH by CONACYT (CVU No. 46280) and KAM is
supported, in part, by STFC grant ST/J001546/1.

\bibliography{non_dust}

\end{document}